\documentstyle[twocolumn,amssymb,aps,psfig,float]{revtex}

\begin{document}
\tolerance 50000

\draft

\twocolumn[\hsize\textwidth\columnwidth\hsize\csname @twocolumnfalse\endcsname

\title{Transverse transport in coupled strongly correlated electronic chains}

\author{S. Capponi and D. Poilblanc}
\address{
Laboratoire de Physique Quantique and C.N.R.S. (UMR 5626), \\ 
Universit\'e Paul Sabatier, 31062 Toulouse, France.
}

\maketitle

\begin{abstract}
\begin{center}
\parbox{14cm}{
One--particle interchain hopping in a system of coupled Luttinger liquids is 
investigated by use of exact diagonalizations techniques. Firstly, the two
chain problem of spinless fermions is studied in order to see the behaviour
of the band splitting as a function of the exponent $\alpha$ which
characterizes the $1D$ Luttinger liquid. 
We give numerical evidence that inter-chain coherent hopping (defined by a
non-vanishing splitting) can be 
totally suppressed for $\alpha\sim 0.4$ or even smaller 
$\alpha$ values.
The transverse conductivity is shown to exhibit a strong incoherent part
 even when coherent inter-chain hopping is believed to occur
(at small $\alpha$ values).
Implications for the optical experiments in
quasi-1D organic or high-$T_c$ superconductors is outlined. 
}
\end{center}
\end{abstract}
]

Recently, the study of strongly correlated fermions confined to coupled
chains has received a great deal of interest in particular as a way of
studying the dimensional cross-over from 1D Luttinger-like behaviour to
2D. 

Some time ago, Anderson emphasized the crucial difference between
 in-plane and inter-plane (c-axis) transport observed in copper oxide
superconductors~\cite{Anderson_91}. Indeed, experimentally the transverse
conductivity has a 
completely incoherent frequency dependence~\cite{Cooper}: there seems to be no sizeable
Drude-like term (except in the optimally doped systems) and $\sigma(\omega)$
is a very slowly increasing function of the frequency.
This phenomenon has been interpreted as an
incoherent hopping or as the ``confinement'' of 
the electrons inside the weakly coupled planes~\cite{Clarke_95}. 
However, for coupled Fermi liquids (FL), Landau theory predicts 
coherent transverse hopping and no anomalous transport. 
Therefore, these data have suggested that the ground 
state (GS) of the two-dimensional (2D) plane itself is not of the usual FL 
type. Unfortunately, it has been impossible yet to prove the NFL nature 
in 2D except for unrealistic models.

However,
it is well known that the generic features of correlated 1D electrons 
are not FL-like but rather those of a Luttinger liquid (LL)~\cite{Haldane} and 
the precise nature (asymptotic behaviour, exponents, etc...) of the system 
can be easily controlled.
In addition, quasi-1D systems are realized in nature, and the problem of coupled
chains is of direct relevance there. For instance, in the case of the organic
superconductors of the (TMTSF)$_2$X family~\cite{jerome}, 
also known as the Bechgaard salts,
the high temperature properties are believed to be essentially one-dimensional,
while the low-temperature behaviour is rather two- or three-dimensional. This cross-over is
presumably responsible for the anomalies observed in the temperature dependence
of several quantities (such as the $2k_F$ contribution to the relaxation 
rate~\cite{wzietek}, the ratio of the 
perpendicular conductivity to the parallel one~\cite{jerome}, 
the plasma edge when the electric
field is polarized perpendicular to the chains~\cite{jerome},\ldots) 
as well as for the
insulating behaviour reported for (TMTSF)$_2$ClO$_4$ in
the presence of a strong enough magnetic field~\cite{behnia}. 
Theoretically a lot of work has
already been devoted to that problem but 
several aspects of this cross-over have 
to be understood better, in particular those concerned with the transport 
properties perpendicular to the chains.

Hence, from now on and for sake of simplicity, we shall only deal 
with weakly coupled chains.

The effect of single-particle transverse hopping has previously been studied from a 
renormalization 
group point of view~\cite{Boies}.
Let us recall here that a LL has a different
structure from FL: there are no quasi-particle like excitations but instead
collective modes (charge and spin) with different velocities which lead to
the so-called spin-charge separation; moreover, the density of states $n(k)$
has no step-like structure at the Fermi level but instead a power 
law singularity
$n(k)-n(k_F)\sim|k-k_F|^\alpha$ defining the parameter $\alpha$ which depends on
the intra-chain interaction. 
It turns out that
the hopping $t_\perp$ is a relevant perturbation when
$\alpha\le 1$ \cite{Boies}. 

However, it has been argued that relevance in
that sense was not necessarily a sufficient condition to cause coherent 
motion between
chains. This, e.g., can be seen from the following model~\cite{Anderson_94};
let a system of two separated chains be prepared at time $t=0$ with a
difference of $\Delta N$ particles between the two chains. Then, the 
interchain hopping is turned on and one considers the probability of 
the system of
remaining in its initial state, $P(t)$. 
Coherence or incoherence can then be defined as the presence or absence 
of oscillations in $P(t)$. 

In Ref.~\cite{Anderson_94}, the authors found two regimes for $P(t)$ which
depend entirely on the value of $\alpha$:
the case $\alpha< 1/2$ exhibited coherent motion while $\alpha>1/2$ showed no
signal of coherence (see also discussion in~\cite{Didier2}). 

This paper is devoted to the study of various aspects of interchain coherence
in systems of strongly correlated spinless fermions. A more extended
version of this work can be found in Ref.~\onlinecite{big_paper}.
We shall derive several quantities sensitive to the
coherent/incoherent nature of the hopping transverse to the chains from
exact diagonalizations of small systems by the Lanczos
algorithm~\cite{Lanczos}.

We consider here a model of spinless fermions on a lattice formed
by $m$ chains of length $L$ with a weak
inter-chain hopping:
\begin{eqnarray}
H=-\sum_{j,\beta} (c^\dagger_{j+1,\beta} c^{\phantom{\dagger}}_{j,\beta} &+& {\rm H.c.})
 - t_\perp \sum_{j,\beta} (c^\dagger_{j,\beta+1} c^{\phantom{\dagger}}_{j,\beta}+ {\rm H.c.})\cr
 &+&\sum_{j,\beta,\delta} V(\delta)\, n_{j,\beta}\, n_{j+\delta,\beta}\nonumber
\end{eqnarray}
where $\beta$ labels the chain ($\beta=1,\dots,m$), $j$ is a rung index
($j=1,\dots,L$), 
 $c_{j,\beta}$ is the fermionic operator, and $V(\delta)$ is a repulsive interaction between two fermions at a
distance $\delta$ (the lattice spacing has been set to one).  For
convenience, we choose a
repulsive interaction of the form $V(i)=2V/(i+1)$ for $i\le i_0$, with, more
specifically, $i_0=1,2,3$ which corresponds to an interaction extending up
to first, second and third nearest neighbours (NN) respectively.

In the x and y-directions, we shall use arbitrary boundary conditions (BC) by threading the system
with a magnetic flux $\Phi_x$ and $\Phi_y$ respectively (except for $m=1$ or 2
chains where open BC are used in the y-direction) which is measured
 in unit of the flux quantum
$\Phi_0=hc/e$. 

The motivation to introduce flux is two-fold. Firstly, as proposed by 
Kohn~\cite{Kohn} transport properties of a correlated system can be 
directly measured from the response of the system to a twist in the boundary
condition. Secondly, our ultimate goal is to extract quantities in the 
thermodynamic limit from
finite size scaling analysis. It turns out that a simple way to improve the 
accuracy for a fixed system size is to average over the boundary conditions 
e.g. over $\Phi_y$ and/or $\Phi_x$~\cite{flux,Gros}.

In order to characterize the behaviour of coupled chains, it is required to,
first, compute the parameters of a single chain for the same model.
Indeed, one important issue is to study whether interchain transport is
a universal function of the LL parameters only or whether it depends on the 
details of the model. 
In the case of NN interactions, the hamiltonian (known as the t--V model)
can be mapped onto a spin chain problem by a Jordan-Wigner transformation
and this is exactly soluble by the Bethe ansatz; thus, $\alpha$ is known for 
each filling~\cite{Haldane}. However, for extended interactions in space, 
a numerical investigation is necessary with the help of conformal invariance
identities. It turns out that the exponent $\alpha$ 
can be related to simple physical quantities~\cite{Voit}
which can be easily extracted from standard exact diagonalization results
using the Lanczos algorithm. 

\begin{figure}[htb]
\begin{center}
\mbox{\psfig{figure=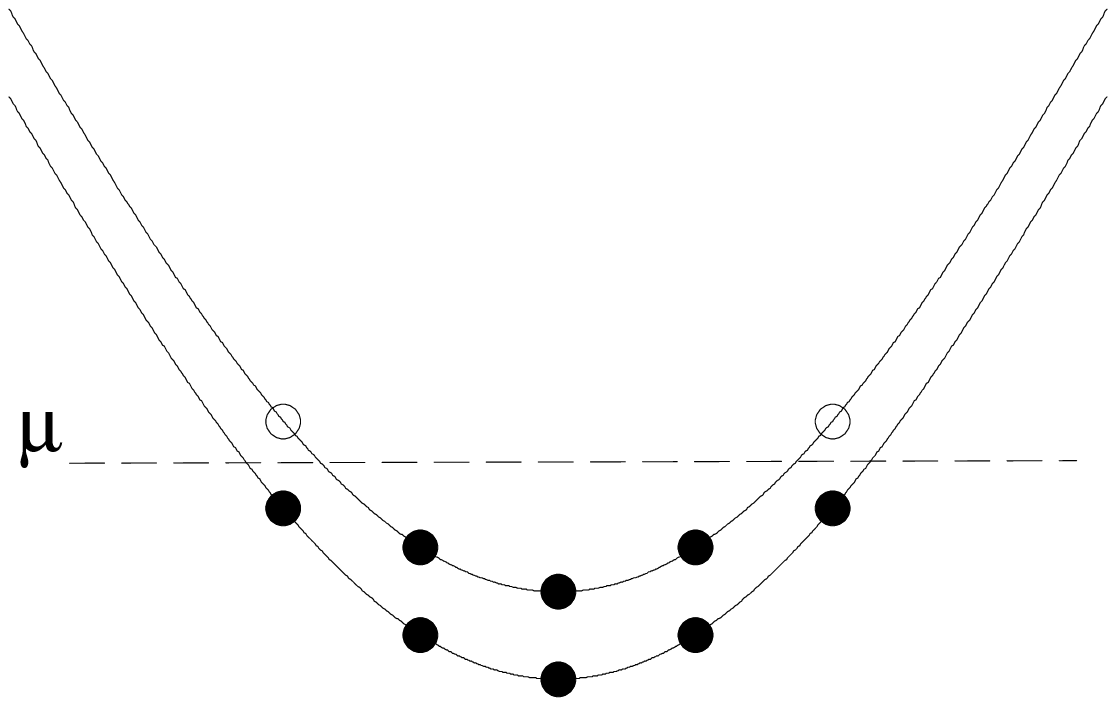,width=4.2cm,angle=0}}
\mbox{\psfig{figure=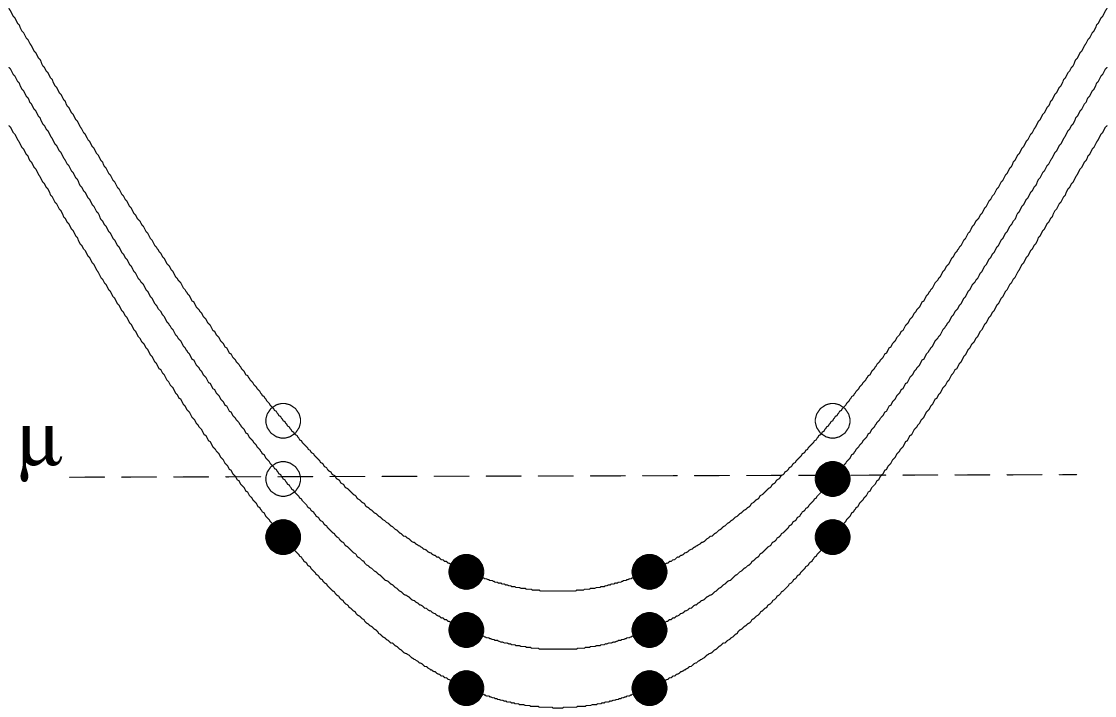,width=4.2cm,angle=0,clip=}}
\end{center}
\caption{Dispersion relation along the chain direction
in the open shell configurations. Full (open)
symbols correspond to occupied (empty) states and 
$\mu$ is the chemical potential.}
\label{nchains}
\end{figure}

The simplest approach to investigate interchain coherence is to 
consider two coupled chains, i.e. a $2\times L$ ladder.
We proceed along the lines of Ref.~\cite{Didier}. 
In the absence of interaction, $t_\perp$ leads to bonding and anti-bonding 
dispersion bands
corresponding to transverse momentum $k_\perp=0$ or $\pi$ respectively,
as seen in Fig.~(\ref{nchains}). 
The splitting $2t_\perp$ between these bands can be viewed as the signature of 
a coherent transverse hopping. 
These bands correspond to a $\delta$-function singularity
in the single particle hole (electron) spectral function 
for $k<k_F$ ($k>k_F$). 

In the case of interacting particles, this $\delta$-function singularity
is replaced by a power law singularity and the elementary excitations
are collective modes. Here, we address the issue of the
influence of the hopping $t_\perp$ on this singularity, in particular we
investigate whether a splitting occurs. 

As can be seen in Fig.(\ref{nchains}), in the open shell configuration,
one can add or remove a particle exactly at the Fermi momentum
which is independent of the system size. 
We thus define the splitting as the difference between 
the electron and hole excitation energies i.e. 
\begin{equation}
\Delta E=E_0(N_e+1,{\bf k}_e)+E_0(N_e-1,{\bf k}_h)-2E_0(N_e).
\end{equation}
where $E_0(N_e)$ is the
reference energy corresponding to the absolute GS of the $N_e=nN$ electron
system. The momenta for the electron and hole excitations are fixed,
${\bf k}_e=(k_F,\pi)$ and ${\bf k}_h=(k_F,0)$.
For V=0, this expression exactly gives the splitting $2t_\perp$ 
for any system size. For $t_\perp=0$ but finite interaction strength,
$\Delta E$ is finite, however it scales to zero in the thermodynamic limit
as expected. We observe too that 
an accurate finite size scaling analysis can be
performed for finite interaction strength and finite $t_\perp$, assuming 
1/L finite size corrections. 

\begin{figure}[htb]
\begin{center}
\psfig{figure=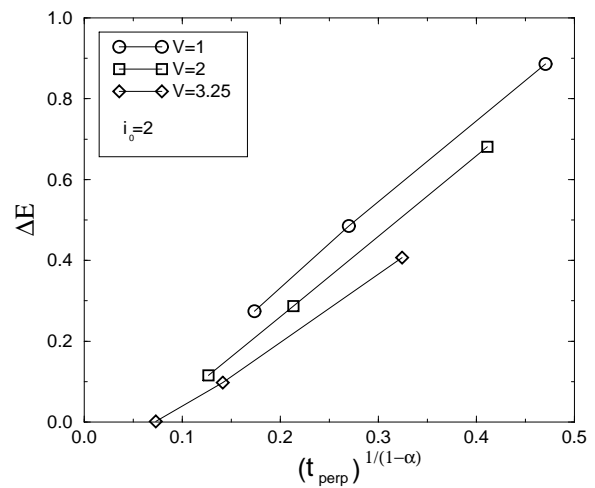,width=\columnwidth,angle=0}
\end{center}
\caption{Extrapolated ($L=\infty$) values of $\Delta E$ vs
$t_\perp^{1/(1-\alpha)}$ in the $i_0=2$ spinless fermion model 
with several values of $V$ (indicated on the figure).
}
\label{Splitting_vsTperp.fig}
\end{figure}

The extrapolated values ($L=\infty$) of $\Delta E$
are plotted as a function of $t_\perp^{1/(1-\alpha)}$ as suggested by the renormalization group
approach~\cite{Boies} 
for various interactions ($i_0=2$ model) in Fig.~(\ref{Splitting_vsTperp.fig}).

A strong reduction of this splitting for increasing $\alpha$ indicates that
intrachain repulsion has a drastic influence in prohibiting
interchain coherent hopping. For large interactions 
like $V=3.25$ ($\alpha\sim 0.38$), there is a critical value 
$t_\perp^*(\alpha)$ of $t_\perp$ below which incoherent transverse
hopping takes place. This corresponds to the case $\alpha > \alpha_0$.
For smaller interaction like $V=1$ ($\alpha\sim 0.08$) our data are
consistent with a coherent behaviour when $t_\perp \rightarrow 0$. Hence, transverse hopping remains coherent in
this case which corresponds to $\alpha < \alpha_0$.
For intermediate interactions like $V=2$ ($\alpha\sim 0.22$),
our data are not conclusive. However, very small values of 
$\alpha_0$ like 0.2 (or even smaller) 
are not inconsistent with our numerical analysis. 

When spin is taken into account, 
the spin-charge separation that
occurs in 1D should suppress even further the coherent transverse hopping.
At a qualitative level, this can be understood from the fact that only 
real electrons can hop from one chain to the next
and this is believed to become more difficult in the presence of spin-charge
separation~\cite{Anderson_91}.
Although dealing with a different filling $n=1/3$ and with particles with
spin, the same qualitative behaviour is found numerically in
Ref.~\cite{Didier}; however, the decrease of the splitting with $\alpha$ is
stronger.


The previous study suggests that the interaction tends to confine the 
electrons within the chains, although no complete confinement 
seems to occur at small $\alpha$ values where $\Delta E\ne 0$. 
A better understanding of this phenomenon can be achieved by investigating
the transport properties along 
the y-axis (inter chain) and more precisely the transverse optical
conductivity which is the linear response of the system to a spatially
uniform, time dependent electric field in the transverse direction.
For such a study, a torus geometry is needed ($m\ge 3$) so that a current
can flow around the loop in the y-direction. 
One of the main advantage of the optical conductivity 
is that it can directly be measured experimentally. 

The real part of the optical conductivity can be written as a sum of two parts,
$ \sigma_{yy}(\omega)=2\pi D_{yy} \,\delta(\omega)+\sigma_{yy}^{reg}(\omega)$
where $2\pi D_{yy}$ is the Drude weight in the y-direction.

Note that in order to check the numerical results, we can use an important
f-sum rule~\cite{Maldague}: the integrated conductivity is proportional to
the mean value of the transverse kinetic energy in the ground state.  

As originally noted by Kohn~\cite{Kohn}, $D_{yy}$
can be obtained from the dependence of the ground state energy $E_0$ on 
$\Phi_y$ as
\begin{equation}
\label{Kohn.eq}
2\pi D_{yy}(\Phi_y)=\frac{m^2}{4\pi}\frac{\partial^2 (E_0/N)}{\partial \Phi_y^2}
\end{equation}
where $N=mL$ is the number of sites. 

The previous quantities have been calculated numerically on finite $3\times L$
lattices using the Lanczos algorithm. 
We have chosen a quarter filled band so that
the extrapolation of results for $L=4,8$ and $12$ is possible.


By choosing adequate boundary conditions along x  open shells can be realized as seen in
Fig.~(\ref{nchains}).
In this case, even for a finite system, the Drude weight and the total kinetic 
energy remains finite down to vanishing $t_\perp$ as in the
non-interacting case. We first consider a fixed value of the flux
$\Phi_y$ such that $\frac{\partial E_0}{\partial \Phi_y} =0$. 
In that case, finite size effects are found to be already weak 
for the two largest cases $L=8$ and $L=12$ and the Drude weight is 
strongly suppressed compared to the $V=0$ case~\cite{note_Convergence}. 

\begin{figure}[htp]
\begin{center}
\psfig{figure=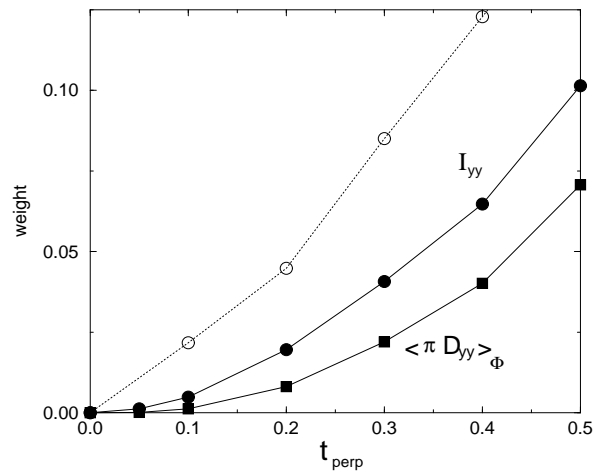,width=\columnwidth,angle=0}
\end{center}
\caption{
$\pi\langle D_{yy}\rangle_{\Phi_y}$ ($\blacksquare$) and $I_{yy}$
({\large $\bullet$}) as a function of $t_\perp$ for a
$3\times 8$ system and $V=2$, $i_0=3$.
For comparison, the $V=0$ case is also displayed as a dotted line.
}
\label{DT.fig}
\end{figure}

In order to mimic the case of many parallel chains, 
an average over $\Phi_y$ is realized~\cite{flux,Gros}
i.e. we calculate $\langle D_{yy} \rangle_{\Phi_y}$ and  
$I_{yy}=\langle \int_0^\infty \sigma_{yy}(\omega) d\omega 
\rangle_{\Phi_y}$ which is obtained by averaging the 
transverse kinetic energy.
We observe that the values obtained for the 
Drude weight by averaging the $3\times 8$ data over $\Phi_y$ 
are very close to the ones obtained on the $3\times 12$ cluster 
at constant flux. Note that, however, if the number of coupled chains is
kept fixed  
 (here $m=3$), even 
in the limit $L\rightarrow\infty$, we expect 
$\langle D_{yy}\rangle_{\Phi_y}\ne D_{yy}(\Phi_y)$. 
In any case, when confinement within each of the individual chains 
starts to occur we do not expect crucial differences between the cases
of 3 or of an infinite number of coupled chains (if only the $t_\perp$ term
couples the chains). 
In Fig.~\ref{DT.fig}, we plot the Drude weight and the total sum rule 
as a function of $t_\perp$ for a moderate interaction.
The behaviour of these two quantities is not 
incompatible with the $t_\perp^2$ law of the non-interacting 
case~\cite{note_Convergence}.
However, the intrachain interaction has drastic effects.
Firstly, the total sum rule is strongly reduced 
compared to $V=0$ (shown as a reference on Fig.~\ref{DT.fig}).
Secondly, it is found that $\pi\langle D_{yy}\rangle_{\Phi_y}$ and $I_{yy}$ 
behave differently.

\begin{figure}[htp]
\begin{center}
\psfig{figure=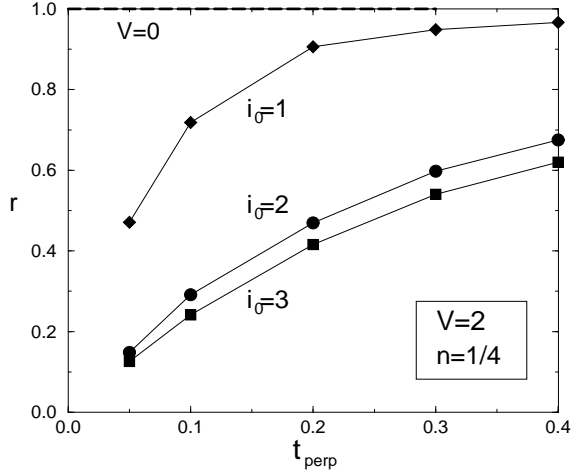,width=\columnwidth,angle=0}
\end{center}
\caption{Ratio of the Drude weight to the total conductivity in the 3x8
system as a function
of $t_\perp$ for NN and 
longer range interaction with the same NN magnitudes $V=2$.}
\label{ratio.fig}
\end{figure}

Let us consider the ratio $r=\pi \langle D_{yy}\rangle_{\Phi_y}/I_{yy}$
 which corresponds to the relative part of the Drude weight
in the total optical conductivity. In Fig.~\ref{ratio.fig}, we note that the ratio $r$ decreases
as $t_\perp$ goes to $0$ and can become rather small. 

These results should have very important consequences on the experimental 
side. Indeed, our results predict, for small $t_\perp$, anomalous transport
perpendicular to the chains even when $D_{yy}$ does not completely 
vanish (if $\alpha<\alpha_0$) since spectral weight is suppressed 
predominantly from the coherent part of the conductivity.
In order to confirm this behaviour, the frequency dependent optical
conductivity has been  calculated directly by use of the Kubo formula,
\begin{equation}
\sigma_{yy}^{reg}(\omega)=\frac{\pi}{N}\sum_{n\ne 0} \frac{|\langle
\phi_0|\,\hat{\j}_y\,|\phi_n\rangle |^2}{E_n-E_0}\, \delta(\omega-(E_n-E_0))
\end{equation}
where $\hat{\j}_y$ is the transverse current operator and the sum runs over
all the excited states.

\begin{figure}[htp]
\begin{center}
\psfig{figure=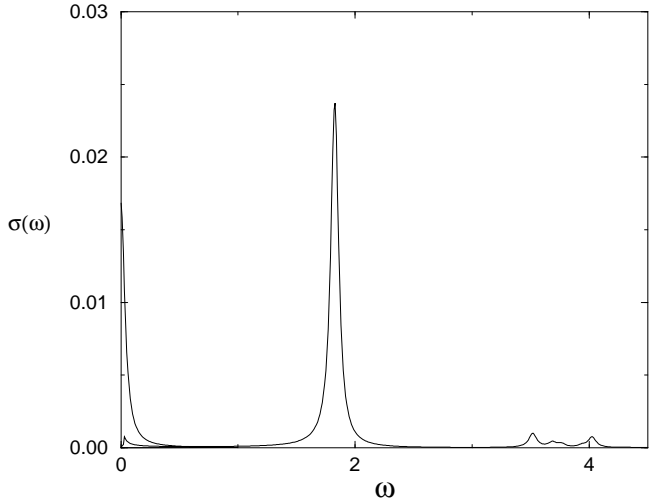,width=\columnwidth,angle=0}
\end{center}
\caption{Transverse optical conductivity vs frequency 
for a $3\times 8$ system at quarter filling with $V=2$, $i_0=3$ and $t_\perp=0.1$.
The Drude $\delta$-function has been represented with the
same small imaginary part $\varepsilon=0.04$.
}
\label{conductivity.fig}
\end{figure}

\begin{figure}[htp]
\begin{center}
\psfig{figure=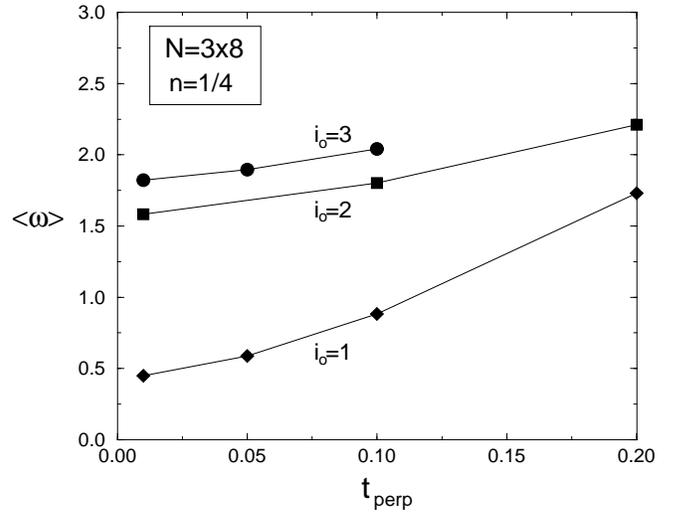,width=\columnwidth,angle=0}
\end{center}
\caption{First moment of the conductivity as a function of $t_\perp$
for $V=2$ and different interaction ranges.}
\label{moment}
\end{figure}

In the free case, the current operator commutes with the Hamiltonian and
therefore, the conductivity only contains a Drude peak.
However, as can be seen on Fig.~\ref{conductivity.fig},
 for finite interaction strengths and in
the $3\times 8$ cluster, a
pronounced structure appears at finite frequency. In order to determine more accurately the position in energy 
of the weight, we have computed the first moment of the distribution,
\begin{equation}
\langle \omega\rangle=\int_{0^+}^\infty
\sigma_{yy}^{reg}(\omega)\,\omega\,d\omega \;/ \int_{0^+}^\infty
\sigma_{yy}^{reg}(\omega)\,d\omega
\end{equation}
which is expected to behave smoothly with the various parameters.
We observe that, once $t_\perp$ is turned on between the chains, weight
immediately appears predominantly at finite frequencies (see
Fig.~\ref{moment}).  
This typical frequency increases with the strength of the interaction.
This is clearly a signature of some form of incoherent perpendicular transport.

In this work, different approaches to interchain coherence have been
investigated. As a first step, we have focused on the energy splitting 
generated by the transverse hopping in the dispersion relation 
of the LL collective modes. By finite size scaling analysis,
we have shown that this splitting was monitored by the LL parameter
$\alpha$.
However, incoherent interchain hopping is found for much smaller
values of $\alpha$ than those predicted by the  RG calculations~\cite{Boies}.
Moreover, in the regime where $t_\perp$ is still relevant 
($\alpha<\alpha_0$), 
the most important results are that (i) the Drude weight and the
total optical sum rule grow 
less rapidly with $t_\perp$ than in the non interacting case 
and (ii) even when the Drude weight remains finite (when $t_\perp$
is relevant), transverse transport 
is predominantly incoherent in the small $t_\perp$ regime.
How small $t_\perp$ needs to be so that this regime is observed 
depends on the strength of the interaction. Typically, for
$\alpha\sim 0.2$, strong suppression of coherent transport occurs
up to $t_\perp\sim 0.15$. 
This phenomenon could explain
the anomalous transport which is observed experimentally.


%
%


\begin{references}
\bibitem{note_Convergence} For a finite system of non-interacting fermions,
the $D_{yy}(t_\perp)$ curve is made of linear segments 
the slope of which increases with $t_\perp$. 
The $t_\perp^2$--law is recovered in the 
thermodynamic limit when the average length of the segments vanishes.
This peculiar behaviour makes finite size scaling analysis of the transverse
conductivity at finite $V$ difficult.
\end{references}
\end{document}